\documentclass[preprints]{Definitions/mdpi} 

\renewcommand{\linenumbers}{}

\firstpage{1} 
\pubvolume{1}
\issuenum{1}
\articlenumber{0}
\pubyear{2023}
\copyrightyear{2020}
\externaleditor{Academic Editor: 
Boris Galperin, Annick Pouquet, and Peter Sullivan
}
\datereceived{} 
\dateaccepted{} 
\datepublished{} 
\hreflink{https://doi.org/} 

\graphicspath{{./fig/}{./png/}}
\usepackage{bm}

\newcommand{\EQ}{\begin{equation}}
\newcommand{\EN}{\end{equation}}
\newcommand{\EQA}{\begin{eqnarray}}
\newcommand{\ENA}{\end{eqnarray}}

\newcommand{\EEq}[1]{Equation~(\ref{#1})}

\newcommand{\Fig}[1]{Figure~\ref{#1}}
\newcommand{\FFig}[1]{Figure~\ref{#1}}

\newcommand{\Tab}[1]{Table~\ref{#1}}

\newcommand{\bra}[1]{\langle #1\rangle}

{}
{}

{}
{}

{}
{}
{}
{}
{}
{}
{}
{}
{}
{}
{}
{}
{}
{}

{}

{}

{}
{}
{}

\newcommand{\xx}{\bm{x}}

\newcommand{\rr}{\bm{r}}

\newcommand{\BB}{\bm{B}}

\newcommand{\JJ}{\bm{J}}

\newcommand{\AAA}{\bm{A}}

\newcommand{\uu}{\bm{u}}

\newcommand{\nab}{{\bm{\nabla}}}

\newcommand{\OO}{\bm{\Omega}}

\newcommand{\SSSS}{\mbox{\boldmath ${\sf S}$} {}}

\newcommand{\ii}{{\rm i}}

\newcommand{\DD}{{\rm D} {}}

\newcommand{\dd}{{\rm d} {}}

\newcommand{\const}{{\rm const}  {}}

\def\Sp{\mbox{\rm Sp}}

\def\Co{\mbox{\rm Co}}

\def\Co{\mbox{\rm Co}}

\def\Lu{\mbox{\rm Lu}}

\def\EEK{{\cal E}_{\rm K}}
\def\EEM{{\cal E}_{\rm M}}

\def\HHM{{\cal H}_{\rm M}}

\def\EK{E_{\rm K}}
\def\EM{E_{\rm M}}

\def\cs{c_{\rm s}}

\def\xiM{\xi_{\rm M}}

\def\EM{E_{\rm M}}

\def\kB{k_{\rm B}}

\def\kB{k_{\rm B}}

\def\Brms{B_{\rm rms}}

\def\urms{u_{\rm rms}}

\def\half{{\textstyle{1\over2}}}

\def\onesixth{{\textstyle{1\over6}}}

\newcommand{\T}{\,{\rm T}}

\newcommand{\s}{\,{\rm s}}

\newcommand{\m}{\,{\rm m}}

\newcommand{\kg}{\,{\rm kg}}

\newcommand{\A}{\,{\rm A}}

\Title{Turbulence with Magnetic Helicity that is Absent on Average}

\TitleCitation{Turbulence with Magnetic Helicity that is Absent on Average}

\Author{Axel Brandenburg $^{1,2,3,4,\dagger,\ddagger}$\orcidA{}, and Gustav Larsson $^{1,2}$\orcidB{}}

\AuthorNames{Axel Brandenburg and Gustav Larsson}

\AuthorCitation{Brandenburg, A.; Larsson, G.}

\address{%
$^{1}$ \quad Nordita, KTH Royal Institute of Technology and Stockholm University, Hannes Alfv\'ens v\"ag 12, 10691 Stockholm, Sweden (\today) \\
$^{2}$ \quad Oskar Klein Centre, Department of Physics, Stockholm University, AlbaNova, 10691 Stockholm, Sweden\\
$^{3}$ \quad School of Natural Sciences and Medicine, Ilia State University, 0194 Tbilisi, Georgia\\
$^{4}$ \quad McWilliams Center for Cosmology and Department of Physics, Carnegie Mellon University, Pittsburgh, Pennsylvania 15213, USA
}

\corres{Correspondence: brandenb@nordita.org}

\abstract{
Magnetic helicity plays a tremendously important role when it is
different from zero on average.
Most notably, it leads to the phenomenon of an inverse cascade.
Here, we consider decaying magnetohydrodynamic (MHD) turbulence as well as
some less common examples of magnetic evolution under the Hall effect
and ambipolar diffusion, as well as cases in which the magnetic field
evolution is constrained by the presence of an asymmetry in the number
density of chiral fermions, whose spin is systematically either aligned
or anti-aligned with its momentum.
In all those cases, there is a new conserved quantity: the Hosking
integral.
We present quantitative scaling results for the magnetic integral scale
as well as the magnetic energy density and its spectrum.
We also compare with cases were a magnetic version of the Saffman
integral is initially finite.
Rotation in MHD turbulence tends to suppress nonlinearity and thereby
also inverse cascading.
Finally, the role of the Hosking and magnetic Saffman integrals in shell
models of turbulence is examined.
}


\begin{document}

\section{Introduction}

This paper is part of a special issue commemorating the work of Jack
Herring.
His scientific career started off with papers on the effect of the solar
wind on the lunar atmosphere in 1959 \cite{H59}.
In 1961, he extended this work to exoplanet atmospheres \cite{H61}.
He also worked on stellar opacities \cite{H63}.
In all those cases, he was very much ahead of its time.
At the time of Parker's prize-winning paper on the discovery of non-static
solutions \citep{Par58}, the physical reality and properties of the
solar wind were still rather unclear and under-appreciated.
Likewise, Herring's work on stellar opacities was well before proper
numerical stellar structure and evolution models became available; the
Henyey method \cite{Henyey+64} (solving a matrix equation instead of using
an iterative shooting method from both ends) became known only in 1964.
Subsequently, Herring turned to hydrodynamic convection and turbulence --
topics that then determined much of his future work.
During his career, he never really worked on magnetic fields or helicity,
but he did interact with people on a daily basis, who were very much
involved in these subjects, both early on \cite{PFL} and also later
during his career \cite{Pouquet+19}.
It is therefore not surprising that this special issue also extends to
topics involving magnetic fields and helicity.

Having helicity in a system usually requires external factors such as
stratification and rotation \citep{Mof78,Par79,KR80}.
In this sense, the absence of helicity may be regarded as the more
generic situation.
It may therefore also seem natural that helicity does not play an
important role when it is absent on average.
This is believed to be the case in hydrodynamic turbulence, but it
changes when magnetic fields are involved.
Although both kinetic and magnetic helicities are ideal invariants,
only the magnetic helicity has a non-ideal dissipation that is slower
than that of the magnetic energy.
By contrast, the dissipation of kinetic helicity is faster than that of
kinetic energy \cite{Matthaeus+Goldstein82, BS05}.
Therefore, in the magnetic case, helicity plays a very important role
in a way that is unknown in the hydrodynamic context.
But is this still true when the net magnetic helicity is actually zero?

The physical situations of interest here include the decay of primordial
magnetic fields in the early Universe during the radiation-dominated era,
when the electric conductivity is high and the initially generated magnetic
field can only decay.
When the plasma is hot enough, the chirality of fermions also plays an
important role, leading to an interplay with magnetic helicity.
Another situation of interest is when only the Hall effect plays a role,
so there are then no fluid motions, but just the flow of electrons.
This is relevant in neutron star crusts, which are solid, so the ions
are immobile. Another application is to the solar wind on scales below the proton
gyroscale, where the ions create a smooth motionless background. The induction equation with just the Hall effect included leads
to interesting decay dynamics---remarkably similar to ordinary
magnetohydrodynamics (MHD). In all those cases, magnetic helicity can play a role even when it vanishes on average. In those cases, magnetic helicity fluctuations may be responsible for
driving an inverse cascade similar to the case of nonvanishing mean magnetic helicity.

Less obvious examples of the dynamics discussed above include the Sun,
because here the magnetic helicity is usually nonvanishing on average \cite{Rust+Kumar96}.
Even in the solar wind, where the conditions resemble those of decaying
turbulence, the magnetic helicity is observed to be nonvanishing on
average and systematically of opposite signs in the northern and southern
hemispheres \cite{BSBG11}.
Near the ecliptic, however, the magnetic helicity fluctuates around
zero \cite{Matthaeus+Goldstein82} and may also be in a state of decay,
so this may be another example where magnetic helicity fluctuations play
an important role.

\section{Nonhelical Turbulence and the Hosking Integral}

In this section, we discuss the Hosking integral and why it is crucial
to understanding nonhelical MHD turbulence with strong magnetic fields.
Unlike the case of weak magnetic fields, when the dynamics is still
controlled by the presence of hydrodynamic effects, we are dealing here
with effects that are specific to the presence of magnetic fields,
albeit with zero average.
We focus on decaying turbulence.

\subsection{Nonhelical Inverse Cascading and Scaling Relations}

Already in 2001, it was noted that, even in the nonhelical case of
a turbulently decaying magnetic field, there is a small amount of
inverse cascading in the sense that for wavenumbers below the peak,
the magnetic energy spectrum rises with time uniformly for all lower $k$
\cite{Christensson+01}.
The actual amount of this rise was small and one could have argued that
it was just because of numerical inaccuracies.
Subsequent simulations \cite{K+13}, however, confirmed such inverse cascading and those
authors discussed the potential interplay between the shallower kinetic energy
spectrum proportional to $k^2$ and the steeper magnetic energy spectrum
proportional to $k^4$.
The qualitative idea was that the shallower velocity spectrum pushes the
magnetic spectrum upward, which then would drive more kinetic energy at
small $k$, and so forth.

The choice of the initial magnetic energy being proportional to $k^4$
is important here.
When such a spectrum was used in the first numerical simulations
\cite{Christensson+01}, the authors made reference to the early work in
Ref.~\cite{Durrer+98}, where causality arguments were put forward.
Nowadays, however, Ref.~\cite{DC03} has become the standard reference
for the choice of an initial $k^4$ spectrum.
Later, it turned out that with a shallower initial $k^2$ spectrum,
no inverse cascading can be found \cite{Reppin+Banerjee17,B+17}.
The reason for this particular aspect will be discussed in more
detail in this paper.

In 2014, the idea of an inverse cascade in the nonhelical case with a
$k^4$ spectrum became really very clear \cite{BKT15}.
This paper was on the arXiv since April 2014, but the paper was published
only in February 2015.
The results were reproduced in the relativistic context in Ref.~\cite{Zrake14}.
Their work was on the arXiv since July 2014 and made reference to the 2015 paper.
The significance of both findings is that it presents early support for
the subsequent discovery of the Hosking integral as a new invariant in
MHD turbulence at large magnetic Reynolds~numbers.

When the Hosking integral was discovered in Ref.~\cite{HS21}, it was
originally called the ``Saffman helicity invariant''.
As already pointed out in Ref.~\cite{ZSB22}, H.\ K.\ Moffatt informed the
community of the fact that this term may be misleading, because the term
`helicity invariant' is reserved for integrals that are chiral in character.
He also recalled that Saffman never considered helicity in his papers.
The term ``magnetic helicity density correlation integral'' may be more
appropriate, but it is rather clumsy.
Following \cite{Scheko22}, where this quantity was called the Hosking integral,
this term continued being used by others \cite{Uchida+22,SB22}.
It should also be noted that `integral' instead of `invariant' is
appropriate since applications to turbulence apply always to finite
Reynolds and Lundquist numbers.
In this connection, it should be emphasized that the Hosking integral
tends to decay with time in a power-law fashion and that the exponent
decreases with increasing Lundquist number $\Lu$ approximately as~$\Lu^{-1/4}$ \cite{ZSB22}.

The energy decay in turbulence is usually characterized by the energy
spectrum $E(k,t)$.
In the following, we sometimes add the subscripts K and M for kinetic and
magnetic energy spectra and other quantities.
We focus here on magnetic energy spectra, $\EM(k,t)$, which are defined
such that $\int\EM(k,t)\,\dd k=\bra{\BB^2}/2\mu_0\equiv\EEM(t)$ is the magnetic
energy, and $\mu_0$ is the vacuum permeability.
The decay can then be parameterized by $\EEM(t)$ and the magnetic integral
scale, which is defined in terms of the magnetic energy spectrum as
\begin{equation}
\xiM(t)=\left.\int_0^\infty k^{-1}\EM(k,t)\,dk\right/
\int_0^\infty\EM(k,t)\,dk.
\label{xirelation2}
\end{equation}
One can always attempt to describe the relations for $\xiM(t)$ and $\EEM(t)$
through power laws.
In addition, the spectrum can evolve underneath an envelope,
\begin{equation}
\EM(k,t)\leq\const\times k^\beta,
\label{Envelope}
\end{equation}
which is in general different from the initial subinertial range spectrum,
$\EM(k,t_0)=\const\times k^\alpha$, where $\alpha$ is the subinertial
range slope.
The three relations for $\xiM(t)$, $\EEM(t)$, and $\EM(k,t)$ can then
be constrained through dimensional arguments once we have a good idea
about the relevant dimensional quantity that governs the decay.

In 2017, the decay of a nonhelical turbulent magnetic field is found to be
described by an exponent $\beta$ that was determined to be between
$\beta=1$ \cite{BK17} and $\beta=2$ \cite{B+17}, but it was unclear
why any of those two possibilities, or any other one, would have to
be expected.
This is what the Hosking integral now explains, namely that $\beta=3/2$.

\begin{figure}[t]
\hspace{-16pt}\includegraphics[width=\columnwidth]{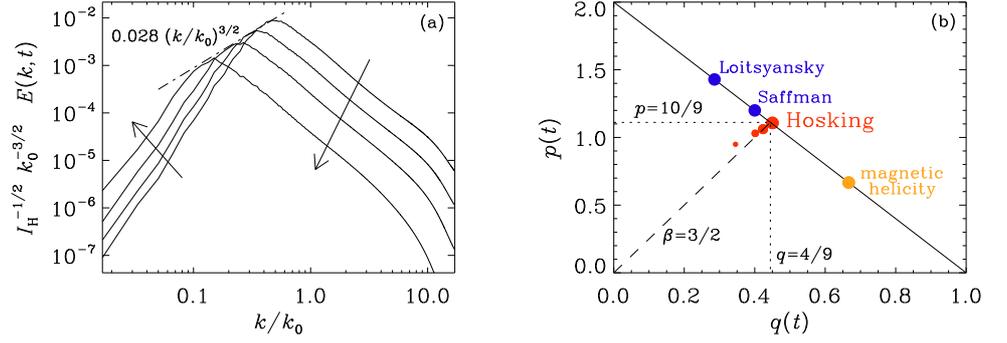}
\caption[]{
(\textbf{a})~magnetic energy spectra, normalized by $I_{\rm H}^{-1/2} k_0^{-3/2}$;
the dashed dotted line shows the envelope $0.028\,(k/k_0)^{3/2}$ under
which the spectrum evolves.
The times are $ck_1t=3$, 7, 17, and 58.
(\textbf{b})~$qp$ diagram showing as red dots the convergence of $p(t)$
versus $q(t)$ toward the Hosking attractor $(q,p)=(4/9,\,10/9)$.
The blue symbols denote the Loitsyansky and Saffman attractors,
respectively, and the orange symbol denotes the magnetic helicity
attractor.
\label{rspec_select_ZSB}}\end{figure}

\FFig{rspec_select_ZSB}a shows magnetic energy spectra at four different
times for a nonhelical magnetically dominated run corresponding to
Run~K60D1bc in Ref.~\cite{ZSB22}.
Here, $k$ is normalized by the initial peak wavenumber $k_0$.
We clearly see that the spectrum exhibits inverse cascading in that
the spectral magnetic energy {\em increases} with time at small $k$,
as indicated by the upward arrow on the left.
The overall energy does of course decay, as indicated by the decline
of the spectral peak and the decrease of spectral energy at large $k$,
as indicated by the downward arrow on the right.

To quantify the temporal changes of $\xiM(t)$ and $\EEM(t)$, it is
convenient to compute the instantaneous scaling exponents \cite{BK17}
\begin{equation}
q(t)=\dd\ln\xiM/\dd\ln t\quad\mbox{and}\quad
p(t)=-\dd\ln\EEM/\dd\ln t;
\end{equation}
see \Fig{rspec_select_ZSB}b.
We see that with time (larger red symbols), the solution evolves toward
the point $(q,p)=(4/9,\,10/9)$, as is also theoretically expected
\cite{ZSB22}.
Although we mainly focus on the case of nonhelical magnetic fields, we
also compare in \Fig{rspec_select_ZSB}b with the expected solution for the fully
helical case (orange), and include solutions for hydrodynamic turbulence
that are governed either by the Loitsyansky or the Saffman integrals.

Before we continue, it is useful to clarify the concept of what we often
refer to as a `governing quantity'.
Take, for example, standard hydrodynamic Kolmogorov turbulence.
Here, the rate of energy transfer per unit mass $\epsilon$ (which is the
rate of energy input and also the rate of energy dissipation) is such
a quantity and the relevant physical scaling laws can be expressed in
terms of powers of $\epsilon$ and other relevant variables such as the
wavenumber $k$ itself.
This then yields for the energy spectrum per unit mass the expression
$E(k)=C_{\rm K}\epsilon^{2/3}k^{-5/3}$, where $C_{\rm K}$ is a
dimensionless coefficient of order unity (the Kolmogorov constant; typically $C_{\rm K}\approx1.6$).
Other such governing quantities include the mean magnetic helicity density
$\bra{h}$ and some other quantities that are crucial to the physics.
They are usually constant or well~conserved.

\subsection{The Loitsyansky and Saffman Integrals in Hydrodynamics}
\label{Loitsyansky}

In the hydrodynamic case, the decay of turbulence can follow different
behaviors depending on the relevant conservation law. (In
practice, conserved quantities are usually not perfectly conserved under
turbulent conditions, and some are better conserved than others.
Which one is the most relevant quantity depends on the relative
conservation properties under different circumstances.)
~One such conserved quantity is the Loitsyansky integral
\cite{D00, D09},
\begin{equation}
I_{\rm L}=-\int\bra{\uu(\xx)\cdot\uu(\xx+\rr)}\,r^2\,\dd^3\rr,
\end{equation}
which is believed to play an important role.
This integral reflects the local conservation of angular momentum and
has dimensions $[I_{\rm L}]=\m^7\s^{-2}$.
If this quantity governs the decay of turbulence, the time dependence
of the growth of the integral scale can be motivated by dimensional
arguments as $\xi(t)\propto I_{\rm L}^a t^b$, where the exponents $a$
and $b$ must be, on dimensional grounds, $a=1/7$ and $b\equiv q=2/7$.
The kinetic energy then obeys $\EEK\propto I_{\rm L}^{2/7} t^{-10/7}$,
i.e., $p=10/7$.
The envelope under which the peak of the spectrum evolves obeys
$\EK(k,t)\leq C_{\rm L} I_{\rm L} k^{4}$.

Another conserved quantity is the Saffman integral,
\begin{equation}
I_{\rm S}=\int\bra{\uu(\xx)\cdot\uu(\xx+\rr)}\,\dd^3\rr,
\end{equation}
which has dimensions $[I_{\rm S}]=\m^5\s^{-2}$.
Similarly, if this quantity governs the decay of turbulence, the time
dependence of $\xi$ must be $\xi(t)\propto I_{\rm S}^a t^b$, where $a=1/5$
and $b\equiv q=2/5$ on dimensional grounds.
The kinetic energy then obeys $\EEK\propto I_{\rm L}^{2/5} t^{-6/5}$,
i.e., $p=6/5$.
The envelope under which the peak of the spectrum evolves obeys
in this case $\EK(k,t)\leq C_{\rm S} I_{\rm S} k^{2}$.

Whether $I_{\rm L}$ or $I_{\rm S}$ determine the decay depends on the
existence of long-range correlations, as can be seen from the
Taylor expansion of the kinetic energy spectrum as \cite{HS21,D00}
\begin{equation}
2\EK(k\to0)
\equiv\Sp(\uu)(k\to0)
=\frac{I_{\rm S}}{2\pi^2}k^2
+\frac{I_{\rm L}}{12\pi^2}k^4+...,
\end{equation}
where an initially non-vanishing Saffman integral automatically implies
a $k^2$ scaling in the subinertial range.
Thus, the decay does depend on the infrared part of the initial kinetic
energy spectrum.
In that case, the slope is the same as that required for the initial
spectrum so that the Saffman integral is indeed nonvanishing.
Furthermore, as pointed out in Ref.~\cite{HS21}, owing to the invariance
of $I_{\rm S}$ and $I_{\rm L}$, both an initial $k^2$ and a $k^4$ spectrum
will remain unchanged.
This implies that there can be no inverse cascading in hydrodynamics.

\subsection{The Magnetic Saffman Integral: Comparison with the Hosking Integral}

As already pointed out in Ref.~\cite{HS21}, the formulation of
Section \ref{Loitsyansky} can also be applied to the magnetic field, except that
there is no reason for the magnetic version of the Loitsyansky integral
to be conserved.
The magnetic Saffman integral (hereafter $I_{\rm SM}$), on the other
hand, might indeed be conserved.
Physically, it would reflect the local conservation of magnetic flux.
Again, when $I_{\rm SM}$ is non-vanishing initially, we expect a quadratic
magnetic energy spectrum, which would also persist at later times.
For a steeper $k^4$ subinertial range magnetic energy spectrum, however,
the magnetic Saffman integral must vanish and the Hosking integral is
then expected to play a dominant role.
It is defined as
\begin{equation}
I_{\rm H}=\int\bra{h(\xx)h(\xx+\rr)}\,\dd^3\rr,
\label{Hintegral}
\end{equation}
where $h=\AAA\cdot\BB$ is
the magnetic helicity density with dimensions
$[h]=[B]^2[x]$.
In ordinary MHD, we can express the magnetic field as an Alfv\'en velocity,
i.e., we write the magnetic field in Alfv\'en units, so $[B]=\m\s^{-1}$.
Therefore, $[h]=[x]^3[t]^{-2}$, and thus
$[I_{\rm H}]=[B]^4[x]^5=[x]^9[t]^{-4}$.
If $I_{\rm H}$ plays a governing role in the decay, we expect therefore
$\xiM(t)\propto I_{\rm H}^{1/9} t^{4/9}$,
$\EEM\propto I_{\rm H}^{2/9} t^{-10/9}$, and
$\EM(k,t)\leq C_{\rm H} I_{\rm H} k^{3/2}$.

The Hosking integral is in general expected to be different from zero
\cite{HS21}.
This automatically implies a quadratic scaling of the helicity variance
spectrum, $\Sp(h)$.
Here, $\Sp(h)=\oint_{4\pi}|\tilde{h}|^2\,k^2\dd\Omega_k/(2\pi L)^3$
denotes the shell-integrated spectrum, a tilde marks a quantity in Fourier
space, and $\Omega_k$ is the solid angle in Fourier space, so that
$\int\Sp(h)\,\dd k=\bra{h^2}$.
The quadratic scaling for a finite Hosking integral follows from the
expansion
\begin{equation}
\Sp(h)(k\to0)
=\frac{I_{\rm H}}{2\pi^2}k^2
+...
\end{equation}
In three dimension, a quadratic spectrum corresponds to white noise.
We also know that the spectrum of a quadratic quantity cannot be more blue
than that of white noise \cite{BB20}, so it seems impossible to have a
helicity variance spectrum whose subinertial range is steeper than $k^2$.

\begin{figure}[t]
\includegraphics[width=\columnwidth]{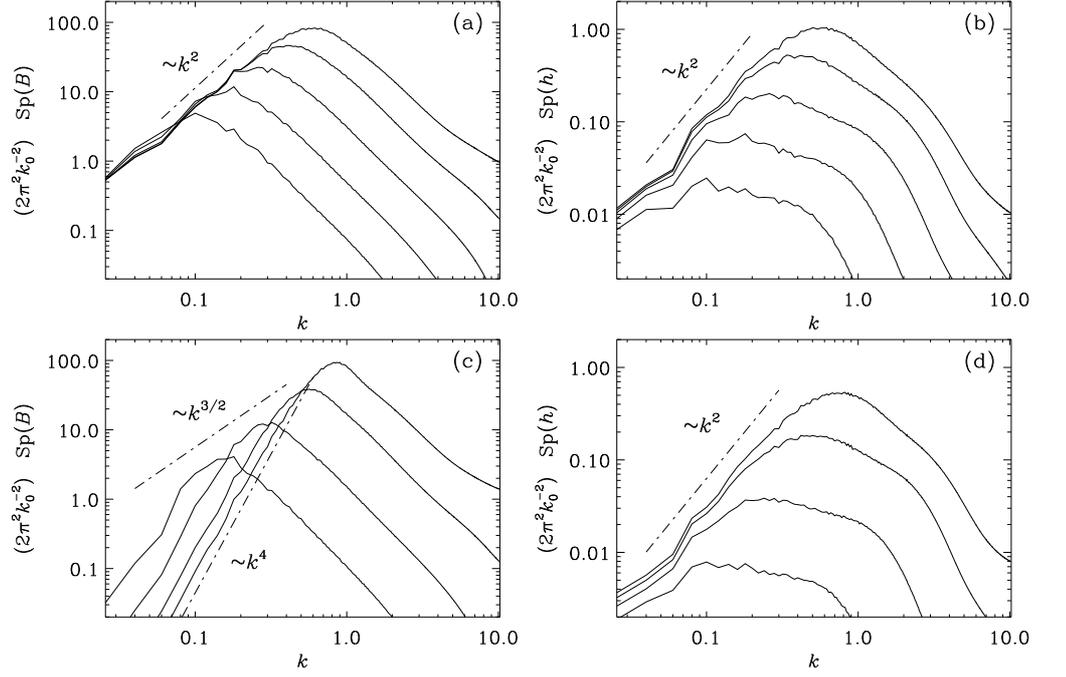}
\caption[]{
Comparison of $\Sp(\BB)$ (\textbf{a},\textbf{c}) and $\Sp(h)$ (\textbf{b},\textbf{d})
for $\alpha=2$ (\textbf{a},\textbf{b}) and $\alpha=4$ (\textbf{c},\textbf{d}).
The dashed-dotted lines indicate $k^{3/2}$ scaling in (\textbf{c}) and $k^2$ scaling otherwise.
\label{rspec_select_comp_uncomp}}\end{figure}

In \Fig{rspec_select_comp_uncomp}, we show magnetic energy and magnetic
helicity variance spectra for initial spectra of the form
\begin{equation}
\EM(k,t_0)=\const\times\frac{k^\alpha}{1+(k/k_0)^{\alpha+5/3}}
\propto
\begin{cases}
k^\alpha & \mbox{for}\;k\ll k_0,\\
k^{-5/3} & \mbox{for}\;k\gg k_0, 
\end{cases}
\end{equation}
for $\alpha=2$ and $\alpha=4$.
We solve the isothermal compressible MHD equations using the
{\sc Pencil Code} \cite{JOSS} with $1024^3$ mesh points.
As expected, and as pointed out \mbox{previously \cite{Reppin+Banerjee17,
B+17}}, there is inverse cascading only for $\alpha=4$, but not
for $\alpha=2$.
Nevertheless, we see that $\Sp(h)$ retains a $k^2$ spectrum at low
wavenumbers in both cases.
This suggests that the Hosking integral is indeed always conserved;
see \Fig{rspec_select_comp_uncomp}b,d.
It may, however, be less dominant than the magnetic Saffman integral.

\begin{figure}[t]
\hspace{-3pt}\includegraphics[width=\columnwidth]{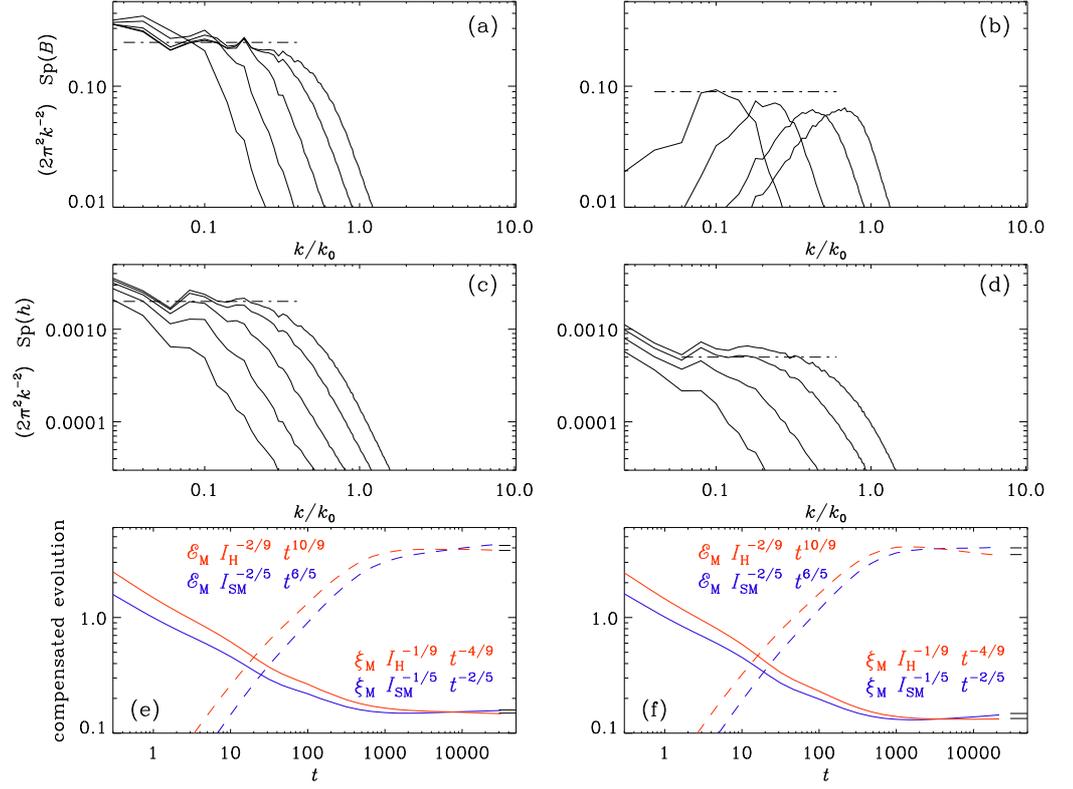}
\caption[]{
Compensated spectra for $\alpha=2$ (\textbf{a},\textbf{c},\textbf{e}) and $\alpha=4$ (\textbf{b},\textbf{d},\textbf{f}).
From (\textbf{a},\textbf{b}), the horizontal dashed-dotted lines indicate that $I_{\rm SM}\approx0.23$
and 0.09, respectively, and
from (\textbf{c},\textbf{d}) they indicate that $I_{\rm H}\approx2\times10^{-3}$
and $5\times10^{-4}$, respectively.
The asymptotic values estimated from (\textbf{e},\textbf{f}) are discussed
in the text and in \Tab{tab1}.
\label{rspec_select_comp}}\end{figure}

To determine the relevant integrals, $I_{\rm H}$ and $I_{\rm SM}$,
it is convenient to plot compensated spectra.
Specifically, to determine $I_{\rm SM}$ and $I_{\rm H}$, we scale
both $\Sp(\BB)$ and $\Sp(h)$ by $2\pi^2/k^2$.
The result is shown in \Fig{rspec_select_comp}.
Thus, in summary, we have
\begin{equation}
\xiM(t)\approx0.16\,I_{\rm SM}^{1/5}t^{2/5},\quad
\EEM(t)\approx4.2\,I_{\rm SM}^{2/5}t^{-6/5},\quad
\EM(k)\approx0.037\,I_{\rm SM} (k/k_0)^{2}.
\label{SaffmanFits}
\end{equation}
If the initial spectrum is not $\propto k^2$, but $\propto k^4$, we have
\begin{equation}
\xiM(t)\approx0.15\,I_{\rm H}^{1/9}t^{4/9},\quad
\EEM(t)\approx3.8\,I_{\rm H}^{2/9}t^{-10/9},\quad
\EM(k)\approx0.025\,I_{\rm H}^{1/2}(k/k_0)^{3/2}.
\label{HoskingFits}
\end{equation}
It is remarkable that the prefactors for the Saffman and Hosking
scalings are very close to each other; see \Tab{tab1} for a summary
of the nondimensional prefactors in the relations
\begin{equation}
\xiM(t)=C_{i}^{(\xi)}\,I_{i}^{\sigma}t^q,\quad
\EEM(t)=C_{i}^{({\cal E})}\,I_{i}^{2\sigma}t^{-p},\quad
\EM(k)=C_{i}^{(E)}\,I_{i}^{(3+\beta)/\sigma}(k/k_0)^{\beta},
\label{GeneralFits}
\end{equation}
where the index $i$ on the integrals $I_i$ and the coefficients
$C_{i}^{(\xi)}$, $C_{i}^{({\cal E})}$, and $C_{i}^{({E})}$ stands for
SM or H for magnetic Saffman and Hosking scalings, respectively, and
$\sigma$ is the exponent with which length enters in $I_{i}$: $\sigma=5$
for the magnetic Saffman integral ($i={\rm SM}$) and $\sigma=9$ for the
Hosking integral ($i={\rm H}$).
Interestingly, while $\beta$ and $p$ can uniquely be related to $q$
via $\beta=2/q-3$ and $p=2(1-q)$ \cite{BK17}, the exponent $\sigma$ is
not uniquely linked to $q$ and we have $\sigma q=2$ for Saffman scaling
and $\sigma q=4$ for Hosking scaling.

The value of $C_{\rm SM}^{(E)}$ only makes sense when $\alpha=\beta=2$,
while that of $C_{\rm H}^{(E)}$ only makes sense when $\alpha=4$
and $\beta=3/2$.
For the other cases, the subinertial range spectrum is not parallel to
$k^\beta$, so $\alpha$ and $\beta$ are said to be incompatible with each
other (see \Tab{tab1}) and the given values of $C_{\rm SM}^{(E)}$ and $C_{\rm H}^{(E)}$
only yield crossings in the middle of the subinertial range.

\begin{specialtable}[t]
\caption{Summary of nondimensional prefactors in the relations
for $\xiM(t)$, $\EEM(t)$, and $\EM(k,t)$.
The numbers in parentheses indicate that the slope $\beta$
is incompatible with the value of $\alpha$.
\label{tab1}}
\begin{tabular}{cccccccc}
\toprule
$\alpha$ & $\beta$ & $C_{\rm SM}^{(\xi)}$ & $C_{\rm H}^{(\xi)}$ & $C_{\rm SM}^{({\cal E})}$ & $C_{\rm H}^{({\cal E})}$ & $C_{\rm SM}^{(E)}$ & $C_{\rm H}^{(E)}$ \\
\midrule
2 &  2  & 0.16 & 0.15 & 4.2 & 3.8 & 0.025 & (0.05) \\
4 & 3/2 & 0.15 & 0.13 & 4.0 & 3.5 & (0.02) & 0.037 \\
\bottomrule
\end{tabular}
\end{specialtable}

We see from \Fig{rspec_select_comp}a that for $\alpha=2$,
the compensated value $(2\pi/k^2)\,\Sp(\BB)\to I_{\rm SM}\approx0.2$.
For $\alpha=4$, on the other hand, we only see a flat
envelope, i.e., $(2\pi/k^2)\,\Sp(\BB)\leq 0.1$, i.e.,
$2\EM(k,t)\leq0.1/(2\pi^2/k^2)\,(k/k_0)^2$.
From \Fig{rspec_select_comp}c,d, 
we see that $(2\pi/k^2)\,\Sp(h)\to I_{\rm H}\approx0.001$
in both cases, i.e., for $\alpha=2$ and $\alpha=4$, respectively.

Given that we now know the values of $I_{\rm SM}$ and $I_{\rm H}$,
we can compensate the time evolutions of $\xiM(t)\propto t^q$ with
$q=2/5=0.4$ and $q=4/9\approx0.44$, and those of $\EEM(t)\propto t^{-p}$
with $p=6/5=1.2$ and $p=10/9\approx1.1$.
The results for the corresponding coefficients in Equation~(\ref{GeneralFits})
are summarized in \Tab{tab1}.

\subsection{The Effect of Rotation}
\label{Rotation}

Rotation suppresses hydrodynamics turbulence.
This is modelled by including the Coriolis force, $-2\OO\times\uu$,
on the right-hand side of the momentum equation, which then reads
\begin{equation}
\frac{\DD\uu}{\DD t}=-\cs^2\nab\ln\rho-2\OO\times\uu
+\frac{1}{\rho}\left[\JJ\times\BB+\nab\cdot(2\rho\nu\SSSS)\right],
\end{equation}
where $\DD/\DD t=\partial/\partial t+\uu\cdot\nab$ is the advective
derivative, $\cs$ is the isothermal sound speed, $\OO$ is the angular
velocity, $\JJ=\nab\times\BB/\mu_0$ is the current density, $\mu_0$ is the
permeability, $\rho$ is the density, $\nu$ is the viscosity, and ${\sf
S}_{ij}=(\partial_i u_j+\partial_j u_i)/2-\delta_{ij}\nab\cdot\uu/3$
are the components of the rate-of-strain tensor.

\begin{figure}[t]
\includegraphics[width=\columnwidth]{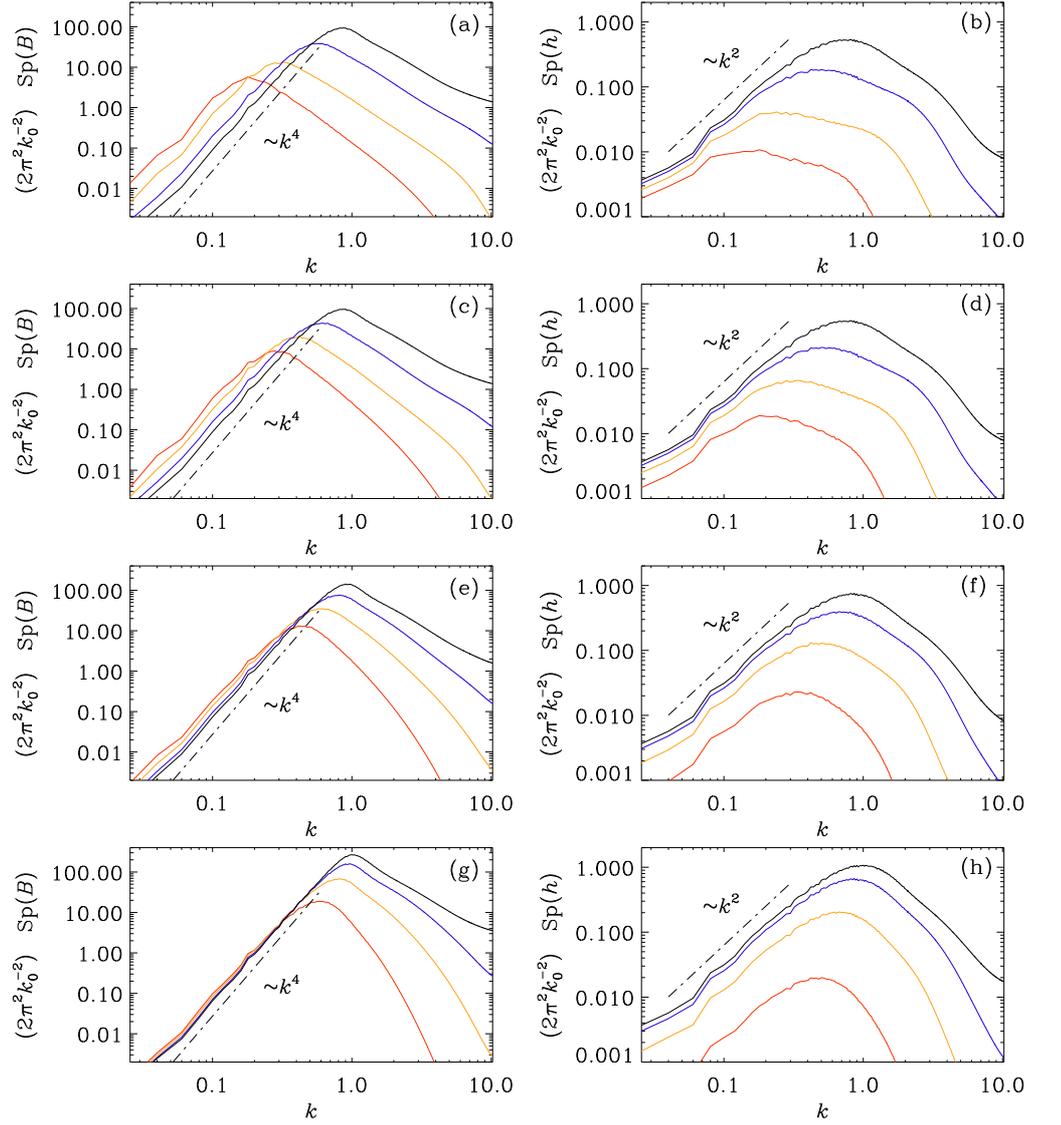}
\caption[]{
$\Sp(\BB)$ (\textbf{a},\textbf{c},\textbf{e},\textbf{g}) and $\Sp(h)$
(\textbf{b},\textbf{d},\textbf{f},\textbf{h}) for $\Omega/\cs k_0$ from $10^{-3}$
(\textbf{a},\textbf{b}: $\Co=1.2$) to $\Omega/\cs k_0=1$ (\textbf{g},\textbf{h}: $\Co=10^4$).
The times are 220 (black), 1000 (blue), 4600 (orange), and 22,000 (red).
\label{rspec_select_comp_rot4}}\end{figure}

In \Fig{rspec_select_comp_rot4}, we show $\Sp(\BB)$ and $\Sp(h)$ for
$\Omega/\cs k_0=10^{-3}$, 0.01, 0.1, and $1$ for runs with $\alpha=4$,
which are otherwise the same as that of \Fig{rspec_select_comp_uncomp}c,d.
We see a clear suppression of inverse cascading already for
$\Omega/\cs k_0=10^{-3}$ and a very strong suppression when this
parameter is unity.
This is caused by the suppression of the turbulent velocity and thereby
of the $\uu\times\BB$ nonlinearity in the induction equation.

To express the angular velocity in a physically more meaningful way,
we note that in the run with $\Omega/\cs k_0=1$, the rms Mach number,
$\urms/\cs$, drops below $10^{-3}$ by the end of the run, while for
$\Omega/\cs k_0=10^{-3}$, it still stays well above $10^{-3}$.
This means that in the latter, the Coriolis number, $\Co\equiv2\Omega/\urms k_0$,
is around unity when rotational suppression becomes appreciable.
At $t=10^4$, the values of $\Co$ are 1.2, 22, 500, and $10^4$ for our
four runs in \Fig{rspec_select_comp_rot4}.

\section{Extensions of the Hosking Idea}

\EEq{Hintegral} is the Hosking integral in its original form.
In the meantime, two further variants of $I_{\rm H}$ have been considered.
One is where $h$ has been replaced by
$h_{\rm tot}=\AAA\cdot\BB+2\mu_5/\lambda$, where $\mu_5$ is the chiral
chemical potential (here in units of an inverse length) and $\lambda$
is a coefficient that quantifies the coupling between fermions and
electromagnetic fields.
The case $\bra{h_{\rm tot}}=0$ has been studied recently in
Ref.~\cite{BKS23}.
Another variant of the Hosking integral is that in the case where the
magnetic field is controlled by the electromagnetic induction from the
Hall effect, which we discuss next.

\subsection{Hall Effect}
\label{HallEffect}

In neutron star crusts, the ions are immobile and the current is only
carried by electrons with the velocity $\uu_{\rm e}=-\JJ/en_{\rm e}$, where
$e$ is the electric charge and $n_{\rm e}$ is the electron density.
As alluded to in the introduction, a similar situation occurs in the solar
wind on scales below the proton gyroscale, where the ions constitute a
smooth background \cite{Cho11}.
The induction equation with the induction from $\uu_{\rm e}\times\BB$
therefore takes the form \cite{GR92}
\begin{equation}
\frac{\partial\BB}{\partial t}=\nab\times\left(
-\frac{1}{en_{\rm e}}\JJ\times\BB-\eta\mu_0\JJ\right),
\label{Hall}
\end{equation}
where $\eta$ is the magnetic field diffusivity.
In the presence of magnetic helicity, one finds an inverse cascade
with an overall decay of the magnetic field and a growth of spectral
magnetic energy at small wavenumbers below that of the peak of the
spectrum \cite{Cho11}.
In the present paper, however, the focus is on the nonhelical case,
which was already considered in B20, but understood mathematically only
later \cite{B23}.

In this context, it is important to note that the natural dimensions
of the magnetic field here are no longer $\m\s^{-1}$, but $\m^2\s^{-1}$.
This was already emphasized in Ref.~\cite{B20}, who used
$e=1.6\times10^{-19}\A\s$, $\mu_0=4\pi\times10^{-7}\T\m\A^{-1}$, and
$n_{\rm e}\approx2.5\times10^{40}\m^{-3}$ for neutron star crusts, so we
have $en_{\rm e}\mu_0\approx5\times10^{15}\T\s\m^{-2}$, and therefore
\begin{equation}
\frac{B}{en_{\rm e}\mu_0}=\frac{B}{5\times10^{15}\T}\,\frac{\m^2}{\s},
\label{DimB}
\end{equation}
which is why we say $B$ has dimensions of $\m^2\s^{-1}$ in the Hall cascade.
This modifies all the dimensional arguments related to $\BB$
correspondingly.
In particular, the units of the magnetic helicity are $[h]=\m^5\s^{-2}$
and those of energy spectra are also $\m^5\s^{-2}$.
Therefore, one has $q=p=2/5$.
This scaling was confirmed in Ref.~\cite{B20}.

\begin{figure}[t]
\includegraphics[width=\columnwidth]{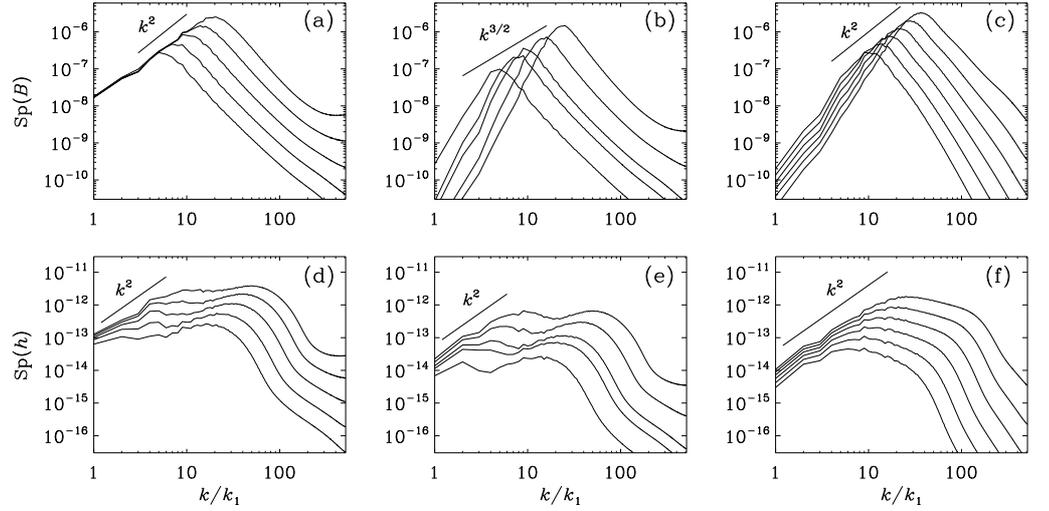}
\caption[]{
$\Sp(\BB)$ (\textbf{a}--\textbf{c}) and $\Sp(h)$ (\textbf{d}--\textbf{f}) for Hall dynamics with
$\alpha=2$ (\textbf{a},\textbf{d}) and $\alpha=4$ (\textbf{b},\textbf{e}), and for ambipolar
diffusion with $\alpha=4$ (\textbf{c},\textbf{f}).
Note the presence of inverse cascading for $\alpha=4$ in panels
(\textbf{b},\textbf{c}), although $\Sp(h)$ changes at $k/k_0\ll1$ in all cases.
The straight lines indicate $k^{3/2}$ scaling in (\textbf{b}) and $k^2$ scaling otherwise.
\label{rspec_select_comp_AD}}\end{figure}

In the nonhelical case, the modified Hosking integral has dimensions
$m^{13}\s^{-4}$, and therefore $q=4/13$ (instead of $4/9$ in MHD).
Furthermore, $p=10/13$ (instead of $10/9$ in MHD), but still $\beta=3/2$
(just like in MHD).
While such a scaling was already seen in the original simulations of
Ref.~\cite{B20}, the work in Ref.~\cite{B23} showed that the modified
Hosking integral is indeed conserved.
In \Fig{rspec_select_comp_AD}a,b,d,e, we show that, also
for Hall dynamics, the Saffman scaling is obeyed for $\alpha=2$, while
Hosking scaling is obeyed for $\alpha=4$.

It should be noted that earlier work on the Hall cascade focussed on
the concept of Whistler turbulence \cite{Gary+08}, where the Whistler
time $t_{\rm w}$ was identified as the governing timescale \cite{Cho11}.
The definition of $t_{\rm w}$ in Ref.~\cite{Cho11} involved the electron
plasma frequency and the electron gyrofrequency such that the electron
mass drops out.
Therefore, $t_{\rm w}$ can more easily be written as
\begin{equation}
t_{\rm w}=\frac{L^2}{\Brms/en_{\rm e}\mu_0},
\end{equation}
which is just the characteristic time based on the magnetic field
expressed in units of a diffusivity; see Equation (\ref{DimB}).
Earlier interpretations in terms of Whistler waves \cite{Cho11} seem
therefore artificial and obscured the relevant interpretation of
the magnetic field as a quantity with dimensions of $\m^2\s^{-1}$.

Observationally, the inverse cascade with helicity may be more easily
accessible, and this has already been tried in the context of neutron
stars \cite{Sarin+23}.
Measuring inverse cascading in the solar wind is conceptually harder,
because the system is statistically steady and more advanced stages
of decay would only correspond to larger distances from the Sun.
On the other hand, most of the spacecrafts are located near the
ecliptic and might therefore allow better insight into nonhelical
inverse cascading.
Unfortunately, most of the observations focus on the high frequency
range of the spectrum \cite{Smith+06}, while the low frequency range is
dominated by noise, making it nearly impossible to say anything about
inverse cascading to even larger scales.

\subsection{Ambipolar Diffusion}
\label{Ambipolar}

The Hall effect is a two-fluid effect where the two components are the
positive and negative charge carriers.
Another two-fluid effect is ambipolar diffusion where
the charged fluid with positive and negative charge carriers is taken as
one component and neutrals are taken as the other component. 
The governing equation is
\begin{equation}
\frac{\partial\BB}{\partial t}=\nab\times\left(
-\frac{\JJ\times\BB}{\rho_{\rm i}\nu_{\rm in}}\times\BB-\eta\mu_0\JJ\right),
\label{Ambi}
\end{equation}
where $\rho_{\rm i}$ is the ion density and $\nu_{\rm in}$ is the
ion--neutral collision frequency.

Unlike the Hall effect in neutron star crusts, where the magnetic
field is said to have dimensions of $\m^2\s^{-1}$, we can here write
\begin{equation}
\frac{B}{\sqrt{\rho_{\rm i}\mu_0}}=\frac{B}{1.5\times10^{-16}\T}\,\frac{\m}{\s},
\end{equation}
where we used $\rho_{\rm i}=1.7\times10^{-26}\kg\m^{-3}$ for the
interstellar medium with an ionization fraction of $10^{-5}$ and a
neutral density of one proton per cubic centimeter.
This is why we say that with ambipolar diffusion, just like in MHD,
$B$ has dimensions of $\m\s^{-1}$.
For this reason, we also see in \Fig{rspec_select_comp_AD}c,f
qualitatively the same decay behavior as in ordinary~MHD.

\subsection{Chiral MHD}

For chiral MHD, the induction equation attains an extra term under the
curl that leads to a contribution to the electric field proportional
to the product of the magnetic field and a pseudoscalar given by the
chiral chemical potential, expressed here as a wavenumber \cite{R+17}.
\begin{equation}
\mu_5=24\,\alpha_{\rm em}\,(n_{\rm L}-n_{\rm R})\,(\hbar c/\kB T)^2,
\label{mu5-tilde}
\end{equation}
where $\alpha_{\rm em}\approx1/137$ is the fine structure constant,
and $n_{\rm L}$ and $n_{\rm R}$ are the number densities
of left- and right-handed fermions, respectively.
The uncurled induction equation takes then the form
\begin{equation}
\frac{\partial\AAA}{\partial t}=\eta(\mu_5\BB-\mu_0\JJ)+\uu\times\BB,
\quad\JJ=\nab\times\BB/\mu_0.
\label{dAAdt}
\end{equation}
The term $\eta\mu_5\BB$ leads to a growth of the magnetic field for
wavenumbers $k<\mu_5$, just in the same way as in mean-field dynamo
theory \citep{Mof78,Par79,KR80}, but here no mean-field theory is invoked.
The generated magnetic field is fully helical, but the relevant
quantity is now the {\em total} chirality density
\begin{equation}
h_{\rm tot}=\AAA\cdot\BB+2\mu_5/\lambda,
\label{htot}
\end{equation}
and it is its volume average that is conserved, i.e.,
$\bra{h_{\rm tot}}=\const$, provided the boundary conditions
are periodic and/or closed, i.e., perfectly conducting.
As the magnetic field grows, $\mu_5$ decreases.
The rate of this change is proportional to the parameter $\lambda$,
which we take here as an adjustable parameter, but in reality is it
given by an expression involving the temperature; see Equation~(49)
of Ref.~\cite{R+17}.

It is important to point out that the {\em physical} chiral chemical
potential (which has the units of an energy) is sometimes defined differently.
First, the authors of Ref.~\cite{BKS23} used Lorentz-Heaviside units,
which implies another factor of $4\pi$ in the numerator of the conversion
factor (or rather the lack of a $4\pi$ factor in the denominator), and,
second, there is also a factor of $2$ in the denominator, so $\hbar
c/8\alpha_{\rm em}$ instead of $\hbar c/4\alpha_{\rm em}$ for the
conversion factor of \cite{BKS23}, because they defined their physical
chiral chemical potential as half the difference between the physical
right- and left-handed chiral chemical potentials.
In addition, there is a sign difference between Refs.~\cite{R+17} and
\cite{BKS23}, but this affects only the physical chiral chemical potential
and not our equations, where $\mu_5$ has the units of a wavenumber.

It turns out, perhaps not surprisingly, that in this case, when
$\bra{h_{\rm tot}}=0$, the turbulence decays again in such a way
that $q=4/9$ and $p=10/9$ and, again, $\beta=3/2$.
This is just like in ordinary (but nonhelical) MHD.
In this case, however, $\bra{h}\neq\const$, but its modulus
decays $\propto t^{-r}$ in a way that is compatible with the
real-space realizability condition, $|\bra{h}|\leq2\EEM\xiM$,
i.e., $r=p-q=(10-4)/9=2/3$.
This was also confirmed in Ref.~\cite{BKS23}.
This study was then applied to the problem of baryogenesis \cite{BKMSS23},
where one tries to explain the small excess of matter over antimatter
in the Universe, which is referred to as baryon~asymmetry.

The Hosking scaling was confirmed for $\bra{\mu_5}\xiM\gg1$ \cite{BKMSS23},
but in the opposite limit of $\bra{\mu_5}\xiM\ll1$ the Hosking scaling
was no longer obeyed and then both $\bra{\mu_5}$ and $\xiM$ are
believed to be approximately independently conserved \cite{BKS23}.
Trying to understand this more thoroughly must be a goal for future
studies, where one may hope to reach much larger scale separation
between the different relevant wavenumbers in the system, such as the
wavenumber $k_0$ of the peak of the magnetic energy spectrum and the
value of $\mu_5$.

\begin{figure}[t]
\includegraphics[width=\columnwidth]{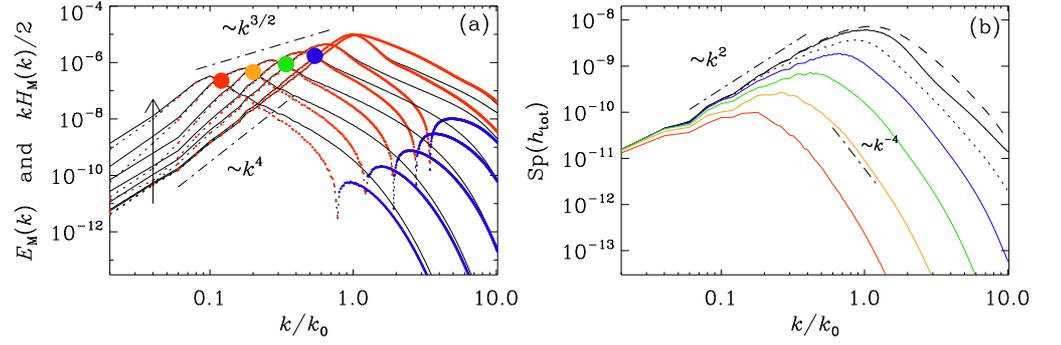}
\caption[]{
(\textbf{a}) Magnetic energy (solid lines) and magnetic helicity spectra (dotted
lines), and (\textbf{b})~magnetic helicity variance spectra for a chiral MHD run
with balanced chirality and an initial $k^4$ spectrum for the magnetic
field.
In (\textbf{a}), positive (negative) magnetic helicities are indicated by small
red (blue) dots.
The four large dots denote the positions of $\xiM^{-1}$.
Their colors are the same as those of the solid lines in (\textbf{b})
and correspond to the times 1,500, 5,000, 15,000, and 50,000.
}\label{rspec_select_hoskM_runI}\end{figure}

\begin{figure}[t]
\includegraphics[width=\columnwidth]{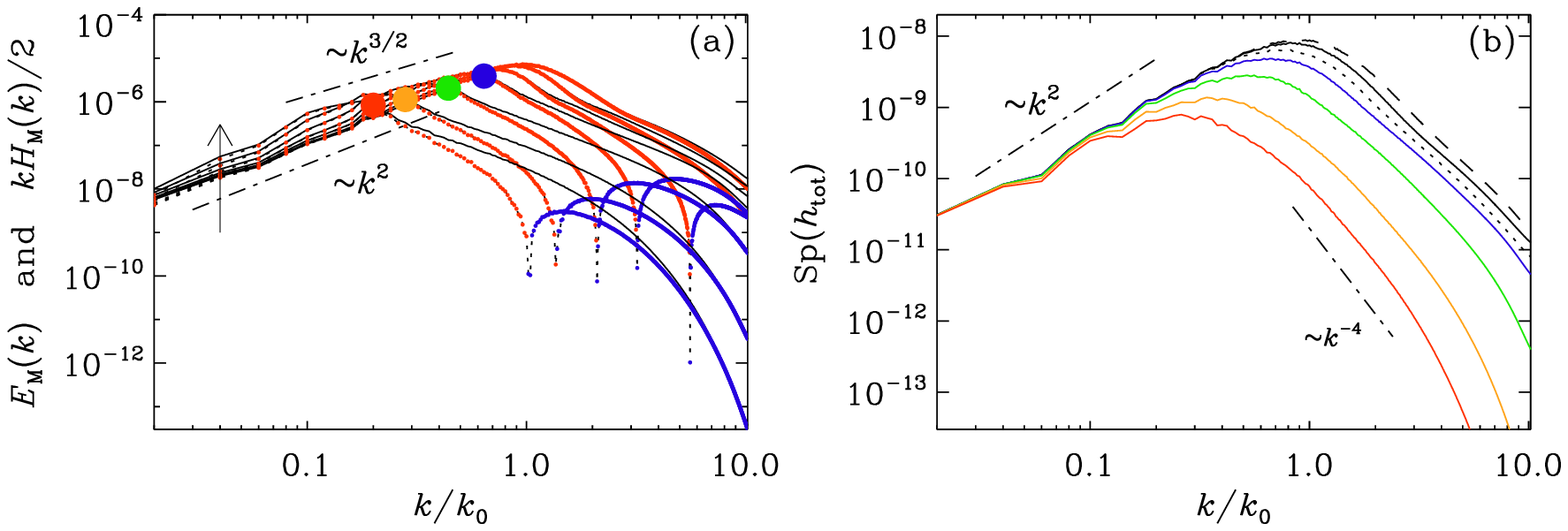}
\caption[]{
Similar to \Fig{rspec_select_hoskM_runI}, but with an initial $k^2$
spectrum.
Note the presence of slight inverse cascading in (\textbf{a}), although
$\Sp(h_{\rm tot})=\const$ at $k/k_0\ll1$ in (\textbf{b}).
The different colors refer to the same times as in \Fig{rspec_select_hoskM_runI}.
\label{rspec_select_hoskM_runC}}\end{figure}

In \Fig{rspec_select_hoskM_runI}, we plot
magnetic energy and magnetic helicity spectra, as well as
magnetic helicity variance spectra for a chiral MHD run
with balanced chirality and an initial $k^4$ spectrum for
the magnetic field.
We see standard inverse cascading with $\beta=3/2$.
Next we compare with the case of an initial $k^2$ spectrum;
see \Fig{rspec_select_hoskM_runC}.
In this case, there is still weak inverse cascading, which
is probably a consequence of the strong contribution from
mean magnetic helicity conservation over extended spatial patches.

Departures from the conservation of the Hosking integral based on
$h_{\rm tot}$ have been seen when $\mu_5<k_0$ \cite{BKMSS23}, but here
we have $\mu_5>k_0$.
To understand more thoroughly the regime where $\mu_5\gg k_0$, we
would need to have much larger numerical resolution.
A possible alternative is to use shell models \cite{Plunian+13}, as will
be discussed next.
However, it is unclear whether such models can capture the relevant
effects related to the Hosking integral or the chiral magnetic effect.

\section{Hosking Integral in Shell Models of Chiral MHD}

Shell models describe turbulence through real or complex scalar variables
on concentric shells in wavenumber space such that certain conservation
laws are obeyed.
In MHD, the relevant conservation laws are those of total chirality,
total (magnetic plus kinetic) energy, and cross helicity.
The Hosking integral describes helicity fluctuations over different scales,
and does not have a direct counterpart at the level of shell models.
However, the scaling properties resulting from its conservation, could
still be manifest in shell models describing the decay of MHD turbulence.

The Hosking integral is particularly important in cases where the mean
chirality vanishes.
It is also conserved otherwise when the mean total chirality is
non-vanishing, but then the conservation of the mean chirality is usually
more important.
It is also important that the magnetic field is strong, because otherwise
the decay properties are dominated by the hydrodynamic turbulent decay.
Our goal here is to investigate the decay of magnetic fields with
vanishing net chirality in chiral MHD using shell models.

In a shell model, we describe the state of the system in shells of
logarithmically spaced wavenumbers $k_n=2^n$, where $n=0$, 1, 2, ...,
$N$ denotes the shell and $N$ is the truncation level.
For $N=30$, for example, we can span ten orders of magnitude
in wavenumber.
In MHD, one usually considers complex variables $B_n$ and $u_n$ for
the magnetic and velocity fields.
The mean magnetic and kinetic energy densities are given by
\begin{equation}
\EEM=\half\sum_{n=0}^N |B_n|^2
\quad\mbox{and}\quad
\EEK=\half\sum_{n=0}^N |u_n|^2.
\label{1}
\end{equation}
In shell models, the fluid density is constant and therefore not indicated
in the definition of the kinetic energy.
Also the permeability factor in the magnetic energy has been omitted.

Magnetic helicity is a signed quantity, i.e., it can be positive or
negative.
How to describe this in a standard shell model is a matter of convention.
One approach is to associate even and odd shells with the decomposition
into positively and negatively polarized modes of the field.
This idea was first developed for the kinetic helicity \cite{Kadanoff+95}.
This then leads to the definition of the magnetic helicity as
\cite{BEO96, FS98, Basu+98}
\begin{equation}
\HHM=\sum_{n=0}^N (-1)^n |B_n|^2/k_n,
\label{2}
\end{equation}
which satisfies the realizability condition
\begin{equation}
k_n|\HHM(k_n)|\leq2\EM(k_n).
\label{3}
\end{equation}

To preserve the preferential growth of positively (negatively), polarized
modes on even (odd) shells, we write
\begin{equation}
\left[\eta k\left(k_n-(-1)^n\mu_5\right)+\frac{\dd}{\dd t}\right]B_n
=\onesixth\ii k_n\,\left[M(u,B)-M(B,u)\right],
\label{4}
\end{equation}
where the $\mu_5$ term leads to a growth of $|b_n|^2$ for even (odd)
values of $n$ when $\mu_5$ is positive (negative), and $M(x,y)$
is a nonlinear functional, where $x$ and $y$ stand for the full
$n$-dependent arrays.
The essence of shell models is to couple only nearest and next-nearest neighbors.
We refer to this model as type~I.
This prescription then leads to
\begin{equation}
M(x,y)= x_{n+1}y_{n+2} + x_{n-1}y_{n+1} + x_{n-2}y_{n-1}
\quad\mbox{(type~I)}.
\label{5}
\end{equation}
As already emphasized in Section \ref{Rotation},
the velocity plays a crucial role in producing an inverse cascade.
It is governed by the Navier-Stokes equation with the Lorentz force
included.
There are then two further quadratic nonlinearities for $u$ and $B$;
see Refs.~\cite{BEO96, FS98, Basu+98} for details.


Another approach to treat helicity is to write the equations
separately for the positively and negatively polarized modes and thus
have evolution equations for $u_n^\pm$ and $B_n^\pm$.
We refer to this model as type~II.
The helicity density can then be written as \cite{BEO97}
\begin{equation}
\HHM=\sum_{n=0}^N \left.\left(|B^+_n|^2-|B^-_n|^2\right)\right/k_n,
\label{2b}
\end{equation}
and the magnetic energy is $\EEM=\sum_{n=0}^N (|B^+_n|^2+|B^-_n|^2)$.
The evolution equations for $B_n^\pm$ take then the form
\begin{equation}
\left[\eta k\left(k_n\mp\mu_5\right)+\frac{\dd}{\dd t}\right]B^\pm_n
=\onesixth\ii k_n\,\left[M_\pm(u,B)-M_\pm(B,u)\right]
\quad\mbox{(type~II)},
\label{4b}
\end{equation}
where \cite{BEO97}
\begin{equation}
M_\pm(x,y)= x^\mp_{n+1}y^\pm_{n+2} + x^\mp_{n-1}y^\mp_{n+1} + x^\pm_{n-2}y^\mp_{n-1}.
\label{5b}
\end{equation}
Note that for the intermediate terms, the signs in the superscripts
are the same, i.e., $u^-_{n-1}B^-_{n+1}$ appear in the evolution of $B^+_n$
and $u^+_{n-1}B^+_{n+1}$ in the evolution of $B^-_n$; see also
Ref.~\cite{Lessinnes+09}, where such models were proposed independently.

\begin{figure}[t]
\includegraphics[width=.49\columnwidth]{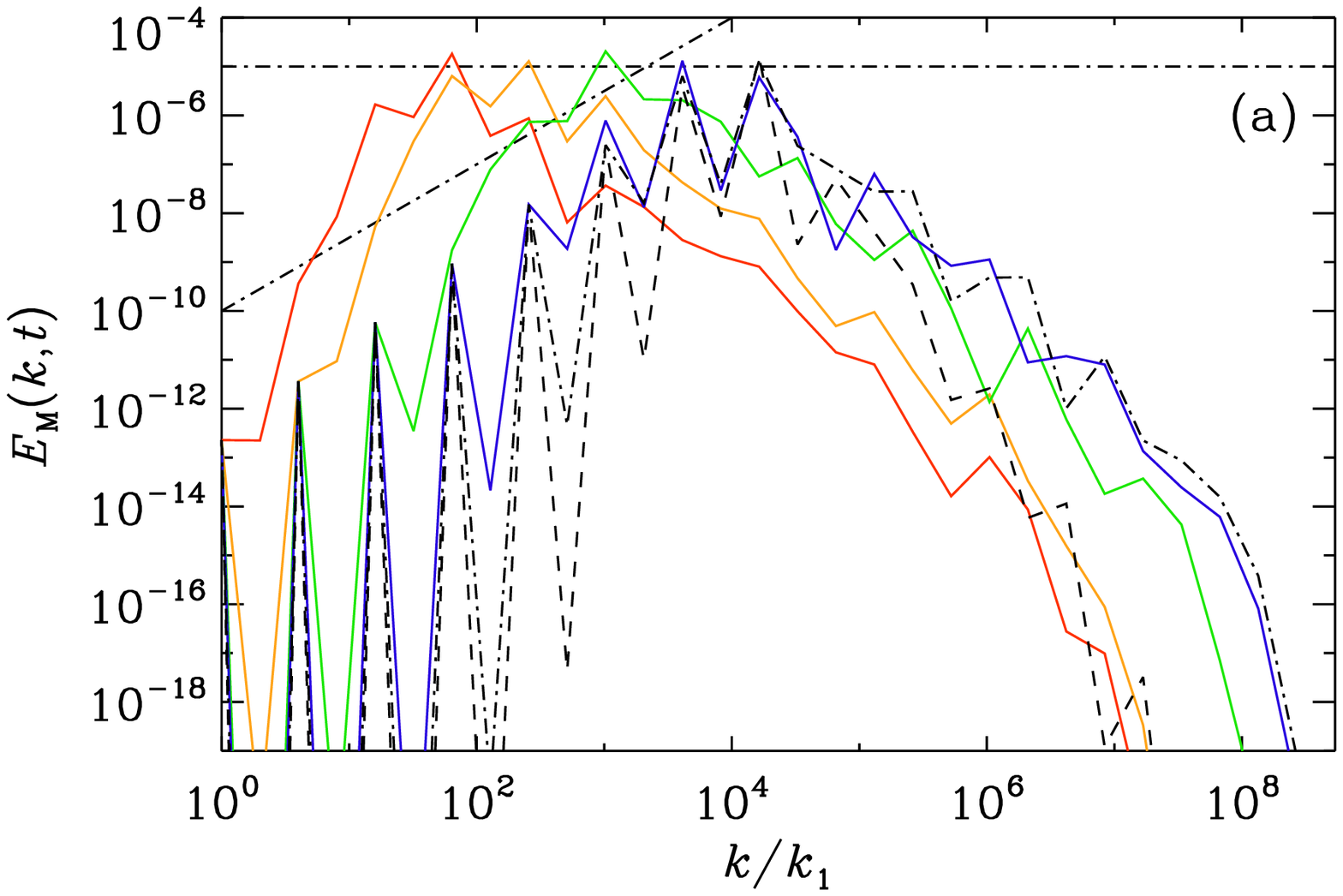}
\includegraphics[width=.49\columnwidth]{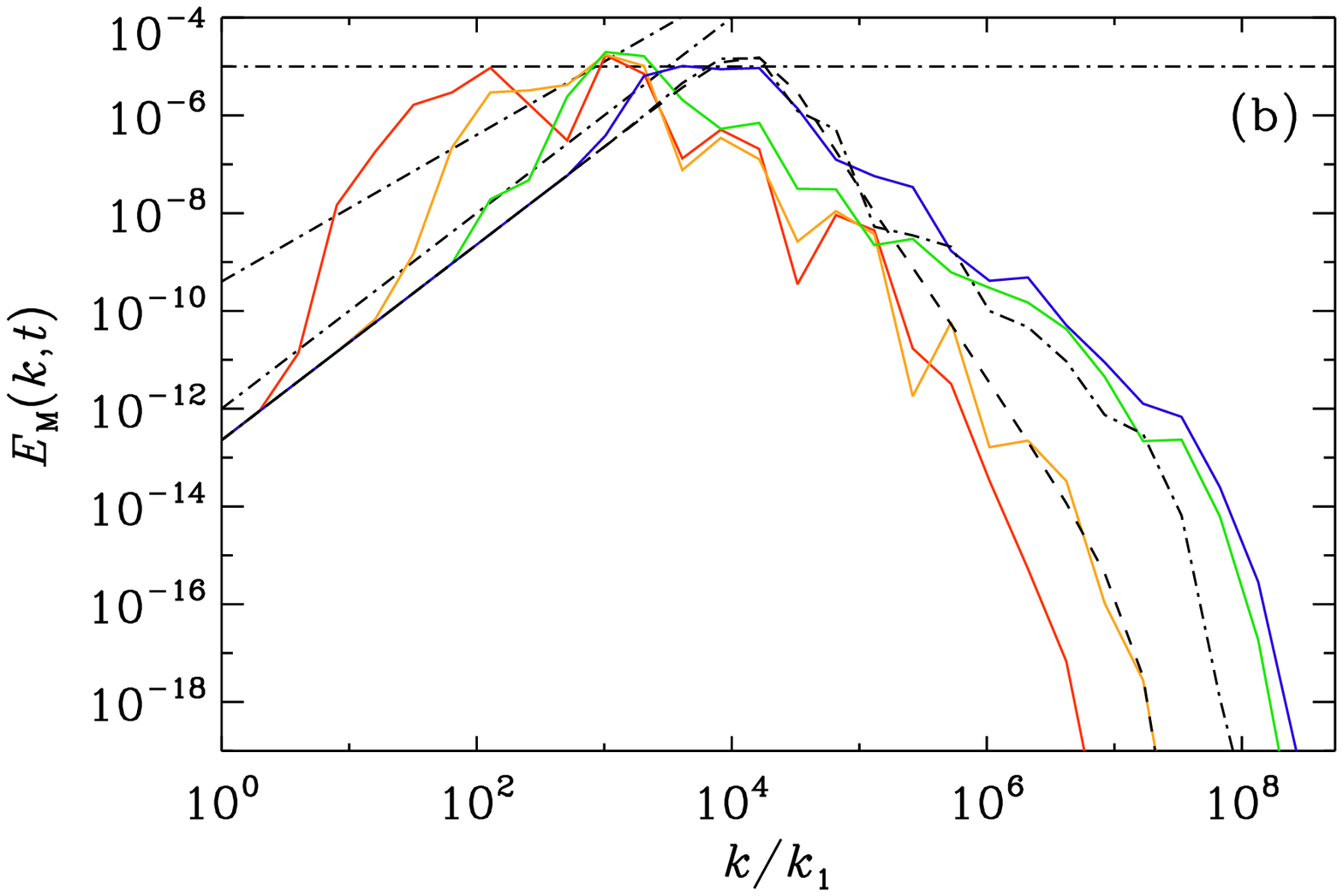}
\caption[]{
Evolution of $\EM(k,t)$ from shell models of (\textbf{a}) type~I and (\textbf{b}) type~II.
The times are 10 (red), 1 (orange), 0.1 (green), 0.01 (blue), and earlier
times are denoted by black lines of different line types.
Note the presence of inverse cascading in both cases.
\label{shell}}\end{figure}

In \Fig{shell}, we present models of types~I and II with $N=30$ shells
using $\lambda=10^{10}$, $k_0=2^{14}=16384\approx1.6\times10^4$,
$\nu=\eta=5\times10^{-11}$, and $\mu_5$ is computed as $\mu_5=-\mu_{\rm M}$,
where $\mu_{\rm M}\equiv\HHM\lambda/2\approx1.8\times10^5$ is the chiral
chemical potential equivalent of the magnetic~helicity.

In all cases, we start with a $k^2$ spectrum, so we expect to see no
inverse cascading.
Looking at the results of \Fig{shell}, however, this does not seem to
be the case.
Our results are still preliminary, but our conclusion so far is that
shell models may not capture the same inverse cascade behavior that we
have found in the direct numerical simulations.
On the other hand, more parameter studies are warranted before one
can draw more firm conclusions.
One must also remember that departures from the conservation of the
Hosking integral have been seen in certain direct numerical simulations
\cite{BKMSS23}.

\section{Conclusions}

In this paper, we have presented a discussion of the Hosking integral
in various contexts in which it has been considered so far: ordinary MHD,
MHD with chiral fermions, as well as just the induction equation --
either with Hall nonlinearity or with ambipolar diffusion nonlinearity.
When the total chirality vanishes (non-chiral case with zero magnetic
helicity or chiral case with finite magnetic helicity balancing the
fermion chirality) it is the correspondingly adapted Hosking integral
that governs the decay of $\EEM(t)\propto t^{-p}$ and the increase of
$\xiM(t)\propto t^q$ with $p=10/9$ and $q=4/9$ for both ordinary MHD
and also just the induction equation with ambipolar diffusion.
When the nonlinearity is given by the Hall effect, on the other hand,
we have $p=10/13$ and $q=4/13$.
The case with chiral fermions is somewhat special, because now the
magnetic field is actually fully helical, but this helicity is balanced
by fermion chirality.
Again, in that case the Hosking integral determines the decay behavior.
However, there is also another decaying quantity: the mean magnetic
helicity density, which is now actually finite and balanced by fermion
chirality.
It is found to decay like $t^{-2/3}$.

In previous work on decaying turbulence, the decay properties
of hydrodynamic and MHD turbulence were motivated by the use of
self-similarity and invariance of the governing equations under
rescaling \cite{Olesen97,BK17}.
This is different in the present work where we have just made use of
dimensional arguments.
Still, the use of invariance under rescaling is necessary to
motivate the equilibrium line $p=2(1-q)$ in the $qp$ diagram in
\Fig{rspec_select_ZSB}b.
It will therefore be interesting to find out whether the existence of
this line could also be motivated by other means.
It probably can, as implied by the derivation of Equation (\ref{GeneralFits}),
and this might reveal a more basic relation to the parameter that
there was called $\sigma$.

An open question is whether the Hosking integral can also play a role
in driven MHD turbulence, for example.
One possibility could be the production of inverse cascade behavior
where magnetic energy grows on wavenumbers below the energy injection
wavenumber.
This could then leads to a turbulent subinertial range scaling of the form
\begin{equation}
\EM(k)\propto I_{\rm H}^a k^b.
\label{6}
\end{equation}
Using dimensional arguments, we would find $3=9a-b$ and $2=4a$ for
balancing the dimensions of length and time, respectively.
Therefore, $a=1/2$ and $b=3/2$.
Thus, $b$ is positive and equal to the Kazantsev slope known in kinematic
nonhelical small-scale dynamos \cite{Kaz68}.
Whether or not there is actually a connection with Kazantsev's small-scale
dynamo theory remains another open question.

In our work we have also examined whether some aspects of the Hosking
integral might be reproducible with shell models.
At the moment, this does not seem to be the case, but this could well be
a consequence of not having performed sufficiently extensive parameter
studies.
Thus, more work might be warranted.

\vspace{6pt}
\authorcontributions{Conceptualization, A.B.\ and G.L.; methodology, A.B.; software, A.B.\ and G.L.
All authors have read and agreed to the published version of the manuscript.}

\funding{This research was funded by Vetenskapsr{\aa}det grant number 2019-04234 and NASA ATP award number 80NSSC22K0825.}




\dataavailability{
The source code used for the simulations of this study,
the {\sc Pencil Code} \citep{JOSS}, is freely available on
\url{https://github.com/pencil-code/} (accessed on 24 May 2023).
The DOI of the code is \url{https://doi.org/10.5281/zenodo.2315093} (accessed on 24 May 2023).
The simulation setups and the corresponding secondary data are available on\\
\url{http://norlx65.nordita.org/~brandenb/projects/Hosking-Shell} (accessed on 24 May 2023).
} 


\conflictsofinterest{The authors declare no conflict of interest.}

\end{paracol}
\reftitle{References}


\externalbibliography{yes}
\bibliography{ref}
\end{document}